\documentstyle[aps,prl,floats,twocolumn,epsf]{revtex}

\begin{document}


\wideabs{
\title{Anisotropic States of Two-Dimensional Electron Systems in 
High Landau Levels: Effect of an In-Plane Magnetic Field}

\author{M.~P. Lilly$^1$, K.~B.~Cooper$^1$, J.~P. Eisenstein$^1$, 
L.~N. Pfeiffer$^2$, and K. W. West$^2$}

\address{$^1$California Institute of Technology, Pasadena CA 91125 \\
         $^2$Bell Laboratories, Lucent Technologies, Murray Hill, NJ 07974}

\maketitle

\begin{abstract}
We report the observation of an acute sensitivity of the anisotropic
longitudinal resistivity of two-dimensional electron systems in
half-filled high Landau levels to the magnitude and orientation of an
in-plane magnetic field. In the third and higher Landau levels, at
filling fractions $\nu=9/2$, 11/2, etc., the in-plane field can lead to a
striking interchange of the ``hard'' and ``easy'' transport directions. In
the second Landau level the normally isotropic resistivity and the weak
$\nu=5/2$ quantized Hall state are destroyed by a large in-plane field and
the transport becomes highly anisotropic. 
\end{abstract}

}

In a recent paper, Lilly {\it et al.}\cite{lilly} reported observations 
of several
anomalies in the low temperature magneto-transport of high quality
two-dimensional electron systems (2DES) when several Landau levels (LLs)
are occupied. These anomalies include the development of large
anisotropies and non-linearities of the longitudinal resistivities (i.e.
$\rho_{xx}$ and $\rho_{yy}$) near half-filling of several spin-resolved
high Landau levels. These effects are observed only at very low temperatures
($T < 150$mK) and when at least three or more Landau levels are
occupied. Taken together, the observations offer strong evidence for new
correlated many-electron states in high Landau levels which are
different than the familiar fractional quantized Hall states found in
the lowest ($N=0$) LL. The fact that qualitatively identical phenomena
are found in several adjacent LLs (having $N \ge 2$) points to a generic
mechanism. In this paper we report on an investigation of the behavior
of these unusual phenomena when a magnetic field component $B_\parallel$
in the plane of the 2D system is added to the perpendicular field
$B_\perp$ generating the Landau quantization. Our results show that the
transport anisotropy can be quite sensitive to the in-plane magnetic
field: in some cases $B_\parallel$ appears to rotate the principal axes
of the anisotropy by $90^\circ$. 

Figure 1 illustrates the anisotropy of the longitudinal resistance seen
in high Landau levels at $T=50$mK. The sample used to obtain these
results is a modulation-doped GaAs/AlGaAs heterojunction containing a
2DES with sheet density of $n_s = 2.7 \times 10^{11}$cm$^{-2}$ and a
mobility of $11 \times 10^6$ cm$^2$/Vs. This structure was grown by
molecular beam epitaxy (MBE) on a $\langle 0 0 1 \rangle$ GaAs
substrate. As the insets suggest, the sample geometry consists of a
square mesa, 2.5mm on a side, etched onto a larger square
chip. The sides of the mesa are rotated $45^\circ$ relative to the
natural cleavage directions ($\langle 1 1 0 \rangle$ and $\langle 1
\overline{1} 0 \rangle$) of GaAs. Eight diffused indium Ohmic contacts
are placed at the corners and midpoints of the sides of the square. For
the data in Fig.~1a, the solid curve corresponds to current flowing
between corner contacts along the diagonal of the square which is
parallel to the $\langle 1 1 0 \rangle$ crystallographic direction while
for the dotted curve the current flow is between corner contacts along
the diagonal parallel to $\langle 1 \overline{1} 0 \rangle$. 
\begin{figure}
\begin{center}
\epsfxsize=3.3in
\epsfclipon
\epsffile[68 255 563 468]{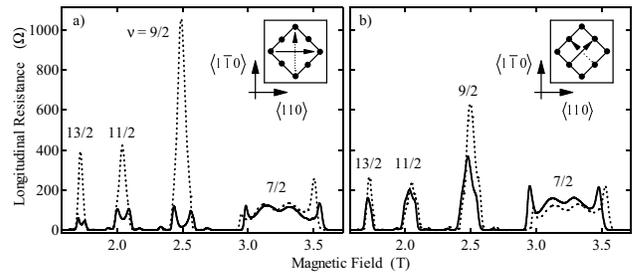}
\end{center}
\caption[figure 1]{Transport anisotropy in high Landau levels in a 
diamond-shaped mesa.  a) current flow along $\langle 1 \overline{1} 0\rangle$
(dotted) and $\langle 1 1 0 \rangle$ (solid). b) current flow along 
$\langle 0 1 0 \rangle$ (dotted) and $\langle 1 0 0 \rangle$ (solid).}
\end{figure}
The inset
to the figures identifies these diagonals. For each trace, the measured
voltage is that between the two midpoint contacts on one side of the
current flow axis. These two orthogonal
resistance measurements yield vastly different results when the 2DES
Fermi level is near half filling of several spin resolved LLs. The
largest anisotropy is observed in the $N=2$ LL near $\nu=9/2$ (where
$\nu = n_s h / e B_\perp$ is the LL filling fraction). At this filling a
deep minimum is seen when the current is driven along $\langle 1 1 0
\rangle$, while a tall peak is found with the current along $\langle 1
\overline{1} 0 \rangle$. For the data shown, the ratio of these
resistances is about 60; in some samples we have found ratios as high
as 3500\cite{lillyreply}. Substantial anisotropies are also seen at 
$\nu=11/2$, 13/2 and
several higher half-odd integers. 
Resistance measurements in square samples can significantly exaggerate
the microscopic resistivity anisotropy\cite{simon}. Even after correcting for these
geometric effects, very large resistivity anisotropies (factor of $\sim
7$ for the $\nu=9/2$ data in Fig.~1) characterize transport in
half-filled high ($N \ge 2$) LLs. This contrasts sharply with the
essentially isotropic transport observed at $\nu=7/2$ and 5/2 in the
$N=1$ LL (shown in Fig.~1) and at $\nu=3/2$ in the $N=0$ lowest LL.
Fig.~1b demonstrates that for currents applied between opposing
midpoints of the square, i.e. along the $\langle 1 0 0 \rangle$ and
$\langle 0 1 0 \rangle$ directions, little anisotropy is seen at any
filling factor. Collectively, these results show that the principal axes
of the transport anisotropy are roughly aligned along the $\langle 1 1 0
\rangle$ and $\langle 1 \overline{1} 0 \rangle$ directions. We emphasize
that the anisotropy axes observed in this 
``diamond'' mesa are precisely the same as was found earlier\cite{lilly}
using a simple cleaved square (from the same parent MBE wafer) having
edges along $\langle 1 1 0 \rangle$ and $\langle 1 \overline{1} 0
\rangle$. This demonstrates that the orientation of the
boundaries of the 2DES does not influence the orientation of the
anisotropy. 

Let us now turn to the effect of an added {\it in-plane} magnetic field
$B_\parallel$. Given the large anisotropies observed near $\nu=9/2$,
11/2, 13/2, etc., it is essential to examine not only the dependence on
the magnitude of $B_\parallel$, but also on its direction. Consequently,
we here report the results of separate experiments with $B_\parallel$
oriented along $\langle 1 1 0 \rangle$ and $\langle 1 \overline{1} 0
\rangle$. Four different cooldowns from room temperature were required,
two for each field orientation. In each case, the magnitude of
$B_\parallel$ was adjusted {\it in situ} at low temperature by tilting
the sample relative to an applied magnetic field. Obviously, the
perpendicular component of the field $B_\perp$ is constant during
studies of a given Landau level filling fraction.

\begin{figure}
\begin{center}
\epsfxsize=3.3in
\epsffile[99 324 495 583]{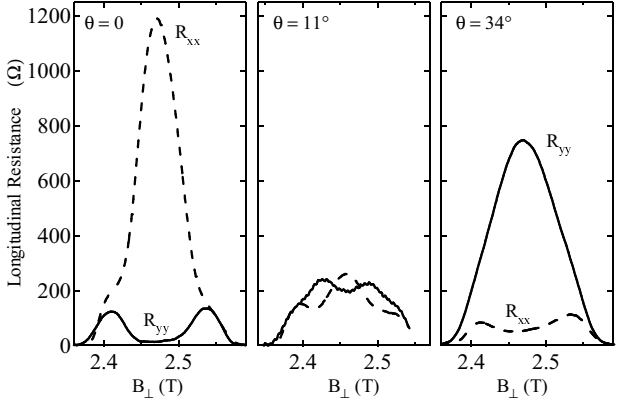}
\end{center}
\caption[figure 2]{Interchange of resistivity anisotropy at $\nu =9/2$
due to an in-plane magnetic field along
$\langle 1 1 0 \rangle$.}
\end{figure}

Figure~2 shows the effect of tilting the sample on the longitudinal
resistance in the vicinity of $\nu=9/2$ when the in-plane magnetic field
lies along $\langle 1 1 0 \rangle$. For these and all subsequent data in
this paper, the sample geometry is a simple $5 \times 5$ mm$^2$ cleaved
square with contacts on the corners and at the midpoints of the sides.
$R_{xx}$ refers to the resistance observed with the current
flowing between midpoint contacts on opposite sides of the square along
a line parallel to the $\langle 1 \overline{1} 0 \rangle$ direction. The
measured voltage difference is that between corner contacts on one side
of the current flow axis. Similarly, $R_{yy}$ refers to the same basic
configuration but rotated by $90^\circ$ so that the average current flow
is along the $\langle 1 1 0 \rangle$ direction. The three panels in the
figure show these resistances at $T=50$mK plotted versus perpendicular
magnetic field $B_\perp$ at tilt angles $\theta=0$, $11^\circ$, and
$34^\circ$. The figure reveals that the large peak in $R_{xx}$ at $\nu=9/2$
seen at $\theta=0$ is rapidly suppressed as the sample is tilted.
Eventually, the peak in $R_{xx}$ is replaced by a minimum. At the same
time, the deep minimum in $R_{yy}$ at $\theta=0$ behaves in essentially
the opposite fashion: it transforms into a tall peak. At around
$\theta=11^\circ$ the two resistance traces are comparable in magnitude
throughout the range $4<\nu<5$. By $\theta=34^\circ$, however, the
resistance at $\nu=9/2$ is again strongly anisotropic, {\it but with
principal axes that are rotated by $90^\circ$ in the 2D plane relative
to their orientation at $\theta=0$.} Thus, the addition of a
sufficiently large in-plane magnetic field directed along the $\langle 1
1 0 \rangle$ changes the ``hard'' transport direction (i.e. high
resistance) at $\nu=9/2$ from $\langle 1 \overline{1} 0 \rangle$ to
$\langle 1 1 0 \rangle$ and the ``easy'' direction (low resistance) from
$\langle 1 1 0 \rangle$ to $\langle 1 \overline{1} 0 \rangle$. 

We find that tilting produces a similar ``interchange effect'' on the
anisotropic resistances at all half odd integer filling from $\nu=9/2$
to 21/2 (and possibly beyond) provided that $B_\parallel$ is directed
along $\langle 1 1 0 \rangle$. This is demonstrated by the left-hand panels of 
Fig.~3 where 
values of $R_{xx}$ and $R_{yy}$ at $\nu=9/2$, 11/2, 13/2,
and 15/2 are plotted versus $B_\parallel$. The figure shows that in each
case the $R_{xx} > R_{yy}$ anisotropy seen at $B_\parallel$=0 (i.e. at
$\theta=0$) gives way to the opposite condition, $R_{xx} < R_{yy}$, at
large $B_\parallel$. Interestingly, the data also show that the cross over point
(where $R_{xx} \approx R_{yy}$) occurs at approximately the same
in-plane field, $B_\parallel \approx 0.5$T, at each filling factor. 

\begin{figure}
\begin{center}
\epsfxsize=3.3in
\epsffile[108 216 468 612]{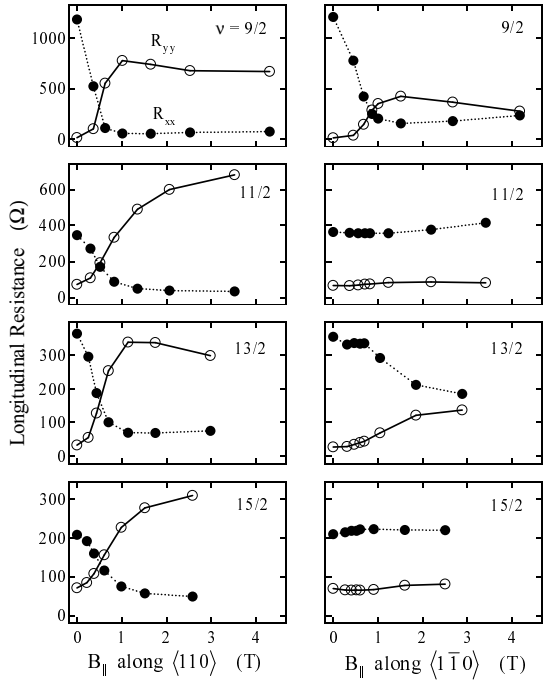}
\end{center}
\caption[figure 3]{Dependence of $R_{xx}$ and $R_{yy}$ at $\nu =9/2$--15/2 on
an in-plane magnetic field at $T = 50$mK. Left panels: $B_\parallel$ along 
$\langle 1 1 0 \rangle$. Right panels: $B_\parallel$ along 
$\langle 1 \overline{1} 0 \rangle$.}
\end{figure}

The right-hand panels of Fig.~3 shows that tilting with $B_\parallel$ directed along the
$\langle 1 \overline{1} 0 \rangle$ direction produces very different
results than with $B_\parallel$ along $\langle 1 1 0 \rangle$. While the
resistances $R_{xx}$ and $R_{yy}$ at $\nu=9/2$ do again interchange, a
larger in-plane field is required and the net transport anisotropy at
large tilt angle is considerably weaker than in the corresponding
experiment with $B_\parallel$ along $\langle 1 1 0 \rangle$. At the
higher half-odd integer fillings the difference between the two
different orientations of $B_\parallel$ is even more striking. At
$\nu=11/2$, for example, applying $B_\parallel$ along $\langle 1
\overline{1} 0 \rangle$ has hardly any effect at all, while for
$B_\parallel$ along $\langle 1 1 0 \rangle$ the resistances interchange
dramatically and lead to an anisotropy ratio at large tilt angles that
exceeds its value at $\theta=0$. This very different behavior between
$\nu=9/2$ and 11/2 is particularly interesting since these two filling
factors belong to the {\it same} $N=2$ orbital LL. In the $N=3$ LL
applying $B_\parallel$ along $\langle 1 \overline{1} 0 \rangle$ does
have a substantial effect on $\nu=13/2$ but it is not nearly so strong
as the interchange observed when $B_\parallel$ is along $\langle 1 1 0
\rangle$. At $\nu=15/2$, like $\nu=11/2$, there is little or no effect with
$B_\parallel$ along $\langle 1 \overline{1} 0 \rangle$. Higher half-odd
integers (e.g. $\nu=17/2$, 19/2, and 21/2) also show virtually no
dependence upon an in-plane magnetic field directed along $\langle 1
\overline{1} 0 \rangle$. 

In some cases, e.g. $\nu=11/2$, 15/2, 19/2, etc., the in-plane magnetic
field appears to exert a ``torque'' on the anisotropy axes. No effect is
seen for $B_\parallel$ along $\langle 1 \overline{1} 0 \rangle$ but a
large effect, which ultimately interchanges the ``easy'' and ``hard''
transport directions, is observed when $B_\parallel$ is along $\langle 1
1 0 \rangle$. This interchange results in the high resistance or
``hard'' transport direction being parallel to $B_\parallel$. It is
tempting to suggest that for these filling factors the anisotropy axes
smoothly rotate as $B_\parallel$ is applied along $\langle 1 1 0
\rangle$. We would then expect to find some intermediate value of
$B_\parallel$ where these axes are rotated by $45^\circ$ relative to
their $B_\parallel$=0 orientation. We searched for this by performing
transport measurments in which the current was driven along $\langle 1 0
0\rangle$ and $\langle 0 1 0 \rangle$ using corner contacts along
diagonals to the square. At no intermediate value of $B_\parallel$ was
a large anisotropy detected along these directions. 

As Lilly {\it et al.}\cite{lilly} and subsequently Du {\it et al.}\cite{du}
observed, the anisotropic transport in
half-filled high Landau levels is a very low temperature ($T<150$mK)
phenomenon. Our data reveal that this largely remains the case in the
presence of the in-plane magnetic field. At very large $B_\parallel$
there is some evidence that the interchanged anisotropy can persist to
higher temperatures than at $\theta=0$ but this is an issue best left to
a future publication. 

Transport near half filling of high Landau levels is, in addition to
anisotropic, substantially non-linear. Lilly {\it et al.}[1] observed
that the large resistivity peak in $R_{xx}$ grows substantially if a dc
current is added to the small ac excitation. Similarly, $R_{yy}$ falls
when the dc current is applied. While the $R_{xx}$ non-linearity is
clearly inconsistent with electron heating, the situation is not as
clear for $R_{yy}$. In any, case we report here that these
non-linearities also appear to interchange in the same way as the
resistances themselves do when an in-plane field is applied. 

The origin of the effects of an in-plane magnetic field on the
anisotropic states of high mobility 2D electron systems at half filling
of highly excited Landau levels is not understood. This is not
surprising since the origin of the large anisotropy in high LLs, even
without any in-plane magnetic field, is itself not understood. Interesting
suggestions, based upon Hartree-Fock calculations, of unidirectional
charge density waves (``stripes'') have been advanced\cite{russians,chalker},
 but quantum
fluctuations are predicted to be severe and to possibly break up the
stripes and produce liquid crystalline behavior\cite{fradkin}. Recent numerical
calculations\cite{kun} have shown that near half filling of high LLs 2D electron
systems become extremely susceptible to periodic potential modulations
having wavelengths comparable to the cyclotron radius. What is clear
from the experiments is that under the appropriate conditions (i.e. very
high sample mobility, very low temperatures, and half filling of the
third and higher LLs) 2D electron systems develop a macroscopic
transport anisotropy. The evidence strongly suggests that this is an
intrinsically many-electron effect. It is also clear that if the
many-electron state has lost full rotational symmetry, some mechanism
must exist for orienting that state over the macroscopic
($\sim$mm) dimensions of our samples. All of the GaAs heterostructure
samples we have studied (11 from 7 different MBE wafers) show, in the
absence of tilting, the same anisotropy axes. The orientation of these
axes appear to be immune to thermal cycling to room temperature,
magnetic field reversal, and changing the geometry of the 2DES (e.g. the
``diamond'' mesa discussed above).
These facts suggest that the anisotropy axes may be determined by
something internal to the MBE-grown GaAs heterostructure, at least in
the absence of an in-plane magnetic field. Possibilities include known
growth instabilities which lead to oblong islands on the sample surface
oriented along the $\langle 1 \overline{1} 0 \rangle$ crystal
direction\cite{orr} and anisotropies in the GaAs bandstructure owing to the
lack of inversion symmetry at the heterointerface\cite{kroemer}. 

An in-plane magnetic field $B_\parallel$ couples to a 2D electron gas
through the electron spin and via the finite thickness of the electron
wavefunction in the direction normal to the 2D plane. In the former
case, $B_\parallel$ adds to the already significant spin Zeeman energy
produced by the perpendicular magnetic field $B_\perp$. The high
sensitivity of the transport coefficients to relatively small
$B_\parallel$ (e.g. at $\nu=9/2$ the 60-fold resistance anisotropy at
$\theta=0$ is wiped out entirely by $\theta \approx 11^\circ$ where
$B_\parallel / B_\perp \approx 0.19$) suggests that small increases in
the spin flip energy are not the most important effect of the in-plane
field. On the other hand, the in-plane field itself leads to mixing of
the different subbands of the heterostructure confinement potential. In
the presence of the perpendicular field, this subband mixing is
accompanied by Landau level mixing. The cyclotron orbits become
anisotropically distorted\cite{cigar} by the in-plane field. Since the resistivity
of the 2D system is highly anisotropic at $B_\parallel = 0$, it is
plausible that such distortions could lead to a sensitivity to the
direction, as well as the magnitude, of the in-plane magnetic field. 

Since the in-plane magnetic field breaks the
rotational invariance of the system, it is
important to examine the dependence of the longitudinal resistance on
the magnitude and direction of the in-plane magnetic field in situations
other than half-odd integer filling factors. For example, in the
Shubnikov-de Haas regime at low perpendicular magnetic fields we find
that the addition of an in-plane field has little effect. As Lilly {\it
et al.} observed[1], the measured resistance $R_{yy}$ typically exceeds
$R_{xx}$ by a factor of about 1.6 in this regime, even though
$B_\parallel=0$. Small anisotropies like this are commonplace in
transport measurements of high mobility 2D electron systems in
GaAs/AlGaAs heterostructures and they in no way obscure the massive low
temperature anisotropies that appear near $\nu=9/2$ and other half-odd
integers. In any case, we find that at $B_\perp \approx 0.5$T the
application of $B_\parallel \approx 0.87$T along either the $\langle 1
\overline{1} 0 \rangle$ or $\langle 1 1 0 \rangle$ axes changes the
resistances, and their ratio $R_{xx} / R_{yy}$, by only about 10\%. This
contrasts sharply with the situation at $\nu=9/2$, 11/2, etc. where, as
Fig.~3a shows, the application of this much in-plane field can changes
this ratio by almost {\it three orders of magnitude.} Similarly, at
high magnetic field, at $\nu=3/2$ in the $N=0$ lowest LL, we also find
relatively little effect of an in-plane field. Applying
$B_\parallel \approx 7.7$T along either $\langle 1 1 0 \rangle$ or 
$\langle 1 \overline{1} 0 \rangle$ yields only a weak increase ($\sim$ 30\%) of the
resistances $R_{xx}$ and $R_{yy}$ at $T=50$mK. In-plane fields of
this magnitude have an enormous impact at $\nu=9/2$. 

\begin{figure}
\begin{center}
\epsfxsize=3.3in
\epsffile[36 295 562 511]{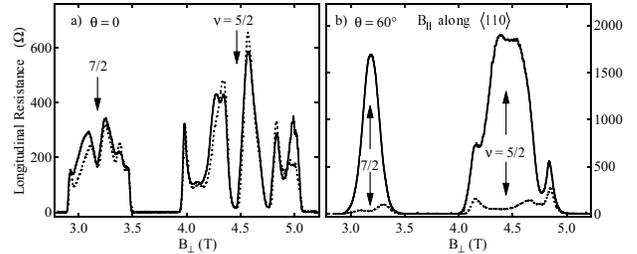}
\end{center}
\caption[figure 4]{Transport at 20mK in the $N=1$ LL. Tilting destroys the
$\nu =5/2$ and 7/2 FQHE states and leaves the transport highly anisotropic.
Current along $\langle 1 1 0 \rangle$ (solid) and $\langle 1
\overline{1} 0 \rangle$ (dotted); $B_\parallel$ along $\langle 1 1 0 \rangle$.
Note the different vertical scales.}
\end{figure}

The remaining case to discuss is the $N=1$ first excited Landau level.
This is the location of the still enigmatic $\nu=5/2$ fractional
quantized Hall effect (FQHE). As we have stressed, in the absence of
tilting, no dramatic transport anisotropies appear near half filling of
either spin branch of this LL (i.e. at $\nu=5/2$ or 7/2). Indeed, the
contrast between the giant low temperature anisotropy of $\nu=9/2$ in
the $N=2$ LL and these essentially isotropic neighbors in the $N=1$
level is one of the most striking results reported by Lilly {\it et al.}
Previous experiments\cite{tilt} have shown that tilting steadily weakens the
$\nu=5/2$ FQHE state. We have corroborated this result using our present
sample by examining the temperature development of $R_{xx}$ and $R_{yy}$
at $\nu=5/2$. On the other hand, as Fig.~4 shows, {\it the resistivity
in the vicinity of $\nu=5/2$ and 7/2 becomes strongly anisotropic as the
sample is tilted.} The orientation of this anisotropy is determined
largely by the direction of the in-plane field: the resistance measured
with current flow parallel to $B_\parallel$ increases steadily upon
tilting while for currents flowing perpendicular to $B_\parallel$ the
resistance changes only slightly. The resulting anisotropies can be
quite significant: at $\nu=5/2$ and $B_\parallel=7.7$T along $\langle 1 1
0 \rangle$ we find $R_{yy} \approx 35R_{xx}$. Since the magnitude of
this $B_\parallel$-induced anisotropy at $\nu=5/2$ and 7/2 is generally
stronger for $B_\parallel$ along $\langle 1 1 0 \rangle$ than $\langle 1
\overline{1} 0 \rangle$ there may also be some sensitivity to the
internal crystal axes.
These new results suggest that the correlation
physics operative at $\nu=9/2$ might be becoming relevant at $\nu=5/2$ and
7/2. In some sense, the $N=1$ LL is straddling the fence between the
FQHE regime in the lowest LL and the spontaneously anisotropic non-FQHE
states which dominate in the $N \ge 2$ LL. If the fragile $\nu=5/2$
state is destroyed by tilting, whether through a spin Zeeman or some
other effect, the real question is: What takes its place at high
$B_\parallel$? Evidently the answer is something anisotropic; perhaps
more akin to the $\nu=9/2$ state in the $N=2$ LL than the composite
fermion liquid which is believed to describe the $\nu=3/2$ state in the
$N=0$ level\cite{review}. 

In summary, we have reported on the effect of an in-plane magnetic field
on transport in high Landau levels. In the third and higher
Landau levels, at filling fractions $\nu=9/2$, 11/2, 13/2, etc., the
recently discovered strongly anisotropic many-electron states near
half-filling are found to be very sensitive to the in-plane field.
Depending upon the direction of the in-plane field, the ``hard'' and
``easy'' transport directions can actually interchange. In the $N=1$
Landau level, we find that the in-plane field not only destroys
the weak $\nu=5/2$ and 7/2 fractional quantized Hall states, but
replaces them with a strongly anisotropic resistivity. 

We are indebted to S. Girvin, A.H. MacDonald, B. Shklovskii, S. Simon,
and K. Yang for valuable discussion and the National Science Foundation
for Grant DMR-9700945.

\end{document}